\documentclass[aps,twocolumn,prd,superscriptaddress]{revtex4-2}
\usepackage{amsfonts}
\usepackage{mathtools}
\usepackage{mathrsfs}
\usepackage{dsfont,amsthm}
\usepackage[hidelinks]{hyperref}
\usepackage{color}
\usepackage{slashed}
\usepackage{amsxtra,mathrsfs,latexsym}
\usepackage{graphicx}
\numberwithin{equation}{section}
\usepackage{subcaption}
\usepackage[justification=raggedright,singlelinecheck=false]{caption}
\usepackage{booktabs}

\newcommand{\diff}{\text{\rm d}}
\newcommand{\imu}{\text{\rm i}}
\newcommand{\euler}{\text{\rm e}}
\newcommand{\Eqref}[1]{Eq.~\eqref{#1}}
\def\sint{\int \!\!\!\!\!\!\!\! \sum_{p}}

\begin{document}

\title{Lee--Yang edge singularities in Nonlocal Nambu--Jona-Lasinio Model}

\author{Zong-shuai Zhang}
\email{zhangzongshuai@stu.pku.edu.cn}
\affiliation{Department of Physics and State Key Laboratory of Nuclear Physics and Technology, Peking University, Beijing 100871, China}

\author{Yi Lu}
\email[]{qwertylou@pku.edu.cn}
\affiliation{Department of Physics and State Key Laboratory of Nuclear Physics and Technology, Peking University, Beijing 100871, China}

\author{Yu-xin Liu}
\email{yxliu@pku.edu.cn}
\affiliation{Department of Physics and State Key Laboratory of Nuclear Physics and Technology, Peking University, Beijing 100871, China}
\affiliation{Center for High Energy Physics, Peking University, 100871 Beijing, China}
\affiliation{Collaborative Innovation Center of Quantum Matter, Beijing 100871, China}

\begin{abstract}

We investigate the QCD phase diagram and the associated Lee--Yang edge singularities using the two-flavor nonlocal Nambu--Jona-Lasinio model extended to complex chemical potential. 
% A close connection is established between the chiral phase transition and the shape of the effective potential as a function of a complex order parameter. 
% A close relation is found between the chiral phase transition and the shape of the effective potential as a function of complex order parameter field, which distinguishes the case between crossover and first-order phase transitions.
There exists a strong correlation between the chiral phase transition and the structure of the effective potential in the complex order parameter plane, serving as a criterion to differentiate crossover from first-order transitions.
Typically, the Lee--Yang edge singularities can be understood as a generalization of the critical end-point (CEP) between crossover and first-order transitions, 
where the positive Nambu phase and the Wigner phase coalesce. 
We further analyze the scaling behavior near the CEP by extracting the critical exponent associated with the Lee--Yang singularities. 
Additionally, we confirm that the extrapolation of the Lee--Yang edge singularity trajectories provides an effective method of determining the CEP location, even at a small real chemical potential. 
This provides a viable method for exploring regions of the QCD phase diagram that remain inaccessible to lattice QCD.

\end{abstract}

\maketitle

\section{Introduction}

In recent decades, considerable attention has been devoted to the study of strong interaction matter at finite temperature and density, due to its relevance to the early universe matter generation and the ultra-relativistic heavy-ion collisions studies. 
The thermodynamic properties of strong interaction matter is embedded in the phase structure of QCD~\cite{ablyazimovChallengesQCDMatter2017,Aarts:2023vsf}, which links to essential features of the interaction, such as the dynamical chiral symmetry breaking, quark/color confinement, and so on.
Over the years, it has been realized that QCD phase structure can be very rich at high baryon chemical potential, typically with a potential critical end-point (CEP)~\cite{Borsanyi:2025dyp,Gao:2020fbl,Gunkel:2021oya,Fu:2019hdw,Hippert:2023bel,Lu:2023mkn,costaPhaseDiagramCritical2010a}, and beyond that the moat regime or inhomogeneous phases~\cite{Fu:2024rto,Motta:2023pks}, color-superconductivity~\cite{Alford:2017ale}, etc. 
Notably, the experimental search of CEP is a hot topic in recent studies, with numerous endeavor put in the studies on specific techniques and relevant observables~\cite{pandavExperimentalSearchQCD2024,Agarwal:2023otg,Pandav:2022xxx,HADES:2019auv,Luo:2017faz,starCollabQCDPhaseDiagram2010,Lu:2025qyf,refId0}. 
It is then crucial to provide estimates not only on the location of critical end-point but also on the properties of critical scaling behaviour through observables. 

In turn, the phase structure at complex chemical potential can also provide valuable information to the properties of CEP mentioned above. 
In 1952, Lee and Yang investigated the distribution of zeros of the partition function in the complex chemical potential plane and their connection to phase transitions~\cite{yangStatisticalTheoryEquations1952, leeStatisticalTheoryEquations1952}. Inspired by their pioneering work, the criticality and analytic behavior of QCD is further investigated by exploring the distribution of the Lee--Yang zeroes and the respective singularities through various theoretical methods~\cite{Dimopoulos:2021vrk,wanLeeYangEdgeSingularities2024,basarQCDCriticalPoint2024,Schmidt:2025ftp,clarkeDeterminationLeeYangEdge2023,clarke2024searchingqcdcriticalendpoint,Mitra:2024czm,Kitazawa:2025cdg,An:2016lni,Johnson:2022cqv,basarUniversalityLeeYangSingularities2021}. 
Nevertheless, systematic investigations of the thermodynamic potential and the (meta)stable phases in the complex order-parameter plane, together with their dynamical evolution as functions of temperature and complex chemical potential, are still scarce. Motivated by this, we aim to perform a detailed analysis of their dynamical behavior at finite temperature and complex chemical potential, from which the distribution of the Lee–Yang edge singularities (LYEs) can be determined.

% (I mean, the shape of potential) 
%
% Quantum Chromodynamics (QCD), the fundamental theory of strong interactions, presents significant theoretical challenges—particularly in the low-energy regime—due to its non-perturbative nature. To address these challenges, various theoretical approaches have been developed, including lattice QCD (LQCD), Dyson-Schwinger equations (DSEs) and others.

% Despite their strengths, these methods have inherent limitations. For example, DSEs constitute an infinite tower of coupled integral equations. Solving them requires adopting a truncation scheme—essentially a method for reducing the infinite set of equations into a closed system\cite{fischerQCDFiniteTemperature2019}\cite{robertsDysonSchwingerEquationsApplication1994}. 

To date, however, theoretical calculations on QCD phase structure are still facing difficulties in the high chemical potential region. 
In particular, lattice QCD simulation as a first-principle approach is still not capable of the case with a large, real chemical potential %for strong interaction matter. 
%$\mu_R$, particularly when $\mu_R/T > 1$ at a given temperature T, 
due to the notorious sign problem~\cite{bazavovFullNonperturbativeQCD2009,detarQCDThermodynamicsLattice2009}. 
To circumvent the sign problem, lattice simulations often employ an imaginary chemical potential and use analytic continuation %has been proposed as a strategy 
to extend the calculation to real chemical potential~\cite{aartsCanStochasticQuantization2009,deforcrandQCDPhaseDiagram2002a}. The analytical continuation from the imaginary chemical potential $\mu_I$ to the real chemical potential $\mu_R$ typically assumes a specific form, such as a Taylor expansion, %e.g. polynomial for the interpolating function, 
which imposes a limiting convergence radius. 

% \cYL{Say functional methods (including DSE~\cite{Ding:2022ows,Fischer:2018sdj,maybe-more} and fRG~\cite{Dupuis:2020fhh}), holographic QCD and effective models are ways for high chem. pot. ... then say we use low energy models here, in particular with an improvement with non-local interaction.}
To go beyond these limitations, continuum QCD approaches have been developed, such as the Dyson--Schwinger equations (DSE)~\cite{fischerQCDFiniteTemperature2019,Bashir:2012fs,robertsDysonSchwingerEquationsApplication1994}, the functional renormalization group (FRG)~\cite{Dupuis:2020fhh}, holographic QCD~\cite{Rougemont:2023gfz,Chen:2022goa,Brodsky:2014yha} and so on, which enable the nonperturbative study on QCD at finite temperature and density including the dynamical properties of strong interaction. 

These advanced approaches typically involve rather sophisticated calculations.  
As a more tractable alternative, low-energy effective models of QCD, such as the Nambu--Jona-Lasinio (NJL) model~\cite{klevanskyNambuJonaLasinioModel1992,hatsudaQCDPhenomenologyBased1994,buballaNJLmodelAnalysisDense2005a}, have been widely employed to investigate the dynamical chiral symmetry breaking(DCSB).
However, the conventional NJL model with a contact interaction and mean-field approximation suffers from various limitations: it is non-renormalizable and requires the introduction of a hard ultraviolet cutoff, which breaks Lorentz invariance. Moreover, the contact interaction fails to reproduce important nonperturbative features of QCD, such as the momentum-dependent quark mass function arising from gluon exchange. 

% To address these issues, nonlocal extensions of the NJL model have been developed\cite{bowlerNonlocalCovariantGeneralisation1994}. By incorporating nonlocal interaction kernels in coordinate or momentum space, the model preserves the chiral symmetry structure while embedding essential aspects of QCD’s low-energy behavior. For instance, the nonlocal form factors can be motivated by Dyson–Schwinger equations\cite{robertsDysonSchwingerEquationsApplication1994}, lattice QCD calculations\cite{parappillyScalingBehaviorQuark2006a}, or the instanton liquid model\cite{schaferInstantonsQCD1998}. Incorporating the dynamics of QCD via a manifest 'running' coupling in terms of a non-local 4-fermion interaction, this model provides a more realistic description of dynamical chiral symmetry breaking. 
% Moreover, the improved ultraviolet behavior of the nonlocal model often eliminates the need for a hard cutoff, thus enhancing theoretical consistency and predictive power.
% \cYL{(for me, the key reason why non-local is superior than contact NJL is as follows:) the dynamics of QCD can be incorporated via a manifest `running' coupling, specifically in terms of a non-local 4-fermion interaction; also, analogy to the gap equation.}

To address these issues, nonlocal extensions of the NJL model have been developed~\cite{bowlerNonlocalCovariantGeneralisation1994}. By introducing nonlocal interaction kernels, also referred to as the form factors in coordinate or momentum space, the nonlocal NJL model preserves the chiral symmetry structure while incorporating essential features of QCD’s low-energy dynamics. 
The nonlocal form factors can be motivated by the results from lattice QCD calculations~\cite{parappillyScalingBehaviorQuark2006a}, DSE or the instanton liquid model~\cite{schaferInstantonsQCD1998}. By embedding QCD dynamics through an explicit momentum-dependent or “running”—coupling in the form of a nonlocal four-fermion interaction, the model offers a more realistic description of the DCSB.
Furthermore, the improved ultraviolet behavior of the nonlocal model typically eliminates the need for a hard cutoff, thereby enhancing the theoretical consistency and predictive capability.
We then carry out our investigation within the framework of the nonlocal NJL model.

This paper is organized as follows. In Sec.~\ref{sec:formalism}, we present the general formalism of the non-local NJL model. In Sec.~\ref{sec:phasediagram}, we investigate the phase diagram at complex chemical potentials, and propose an improved method for identifying the critical endpoint (CEP) and the Lee--Yang edge singularities (LYEs). 
We further analyze the critical behavior by extracting the critical exponent $\beta\delta$ and assess the validity of the LYE-based extrapolation method for locating the CEP.
Finally, Sec.~\ref{sec:conclusion} provides a summary of our findings.

\section{Nonlocal Nambu--Jona-Lasinio Model}\label{sec:formalism}
\subsection{Nonlocal NJL effective action}\label{nonlocal-action} 
The basic idea of the nonlocal NJL model is to approximate QCD interactions with the nonlocal 4-quark interactions.
A scheme based on the features of the instanton liquid model (ILM) has been introduced in Ref. \cite{bowlerNonlocalCovariantGeneralisation1994}, in which the nonlocal form factor is associated to each quark field.
At zero temperature and chemical potential, i.e., in the vacuum, the effective action takes the form:
\begin{equation}\label{Sint}
 \mathcal{S}_E=\int\diff^4 x\,\left[ \bar \psi(x)\left( -i\slashed{\partial} +m_q\right)\psi(x)-\frac{G}{2} J_{\alpha}(x)J_{\alpha}(x)\right]\,,
\end{equation}
here $m_q$ is the current quark mass, and the Euclidean operator $\slashed{\partial}$ is defined as:
\begin{equation*}
    \slashed{\partial} = \gamma_4\frac{\partial}{\partial \tau}+\vec{\gamma}\cdot\vec{\nabla}\,,
\tag{\ref{Sint}a}
\end{equation*}
with $\gamma_4=i\gamma_0, \tau=it$. 
The current $J_{\alpha}(x)$ involving nonlocal interactions is given by:
\begin{equation}
    J_{\alpha}(x)=\int \diff^4 y\int \diff^4 z \, r(y-x)r(x-z)\bar\psi(y)\Gamma_{\alpha}\psi(z)\,,
\end{equation}
here $\Gamma_{\alpha} = (1,i\gamma_5 \vec\tau)$ and $r(x-z)$ is a nonlocal regulator function.

To perform a standard bosonization of the theory for convenience, we can introduce auxiliary fields $\varphi_\alpha(x)=\left(\sigma(x),\vec{\pi}(x)\right)$, where $\sigma$ and $\vec{\pi}$ are chiral partner boson fields representing the scalar-isoscalar and pseudoscalar-isovector mesonic degrees of freedom, respectively. Finally, we obtain the Euclidean generating functional of nonlocal NJL model in momentum space as:
\begin{equation}
		Z=\int\mathscr{D}\sigma\mathscr{D}\vec{\pi}\,\exp[-\mathcal{S}_\text{E}^\text{bos}]\ ,
\end{equation}
with
\begin{equation}\label{Sbos}
		\mathcal{S}_\text{E}^\text{bos}=-\ln\,\det \hat A+\dfrac{1}{2G}\int\dfrac{\diff^4 p}{(2\pi)^4} \,\phi_\alpha^2(p),		
\end{equation}
here,
\begin{align}
    A(p,p') \coloneqq& \langle p|\hat A|p'\rangle \nonumber \\
    =&\left(\slashed{p}+m_q\right)(2\pi)^4 \delta^{(4)}(p-p') \nonumber \\
    &+ r(p)r(p')\varGamma_\alpha\, \phi_{\alpha}(p-p'), \\
    \phi_\alpha(p) =& \int \diff^4x \,\,\euler^{\imu p\cdot x}\varphi_\alpha(x)\,, \\
    r(p) =& \int \diff^4z \,\,\euler^{\imu p\cdot x}\,r(z)\,.
\end{align}
Since $r(p)$ is Lorentz invariant, it should be a function of $p^2$. So we will use the form $r(p^2)$  from now on.

\subsection{Mean Field Approximation and Gap Equation}

In a homogeneous and isotropic vacuum, the translationally invariant vacuum expectation values of the meson fields are given by $\overline\sigma = \langle\sigma\rangle$ and $\overline\pi = \langle\pi\rangle$. From symmetry considerations, the mean values of the pion fields vanish due to their nature as Goldstone bosons.
The mesonic fields can thus be expanded around their mean values as:
\begin{equation}
\sigma(x) = \overline\sigma + \delta\sigma(x), \quad \vec{\pi}(x) = \delta\vec{\pi}(x).
\end{equation}

The bosonized Euclidean action $\mathcal{S}_\text{E}^\text{bos}$ can be expanded in powers of the mesonic fluctuations $\delta\sigma, \delta\vec{\pi}$:
\begin{equation}
\mathcal{S}_\text{E}^\text{bos} = \mathcal{S}_\text{E}^\text{MF} + \mathcal{S}_\text{E}^{(2)} + \dots \label{eq:bosonize}
\end{equation}

The grand canonical thermodynamic potential per four-volume $V^{(4)}$ in the mean-field approximation is:
\begin{align}
\Omega^{\text{MF}}(T,\mu) &= - \frac{T}{V} \, \ln {\cal Z}^{\text{MF}}(T,\mu) \nonumber \\
&= -4 N_\text{c} \int\dfrac{\diff^4 p}{(2\pi)^4} \ln\left[p^2 + M^2(p^2)\right] + \dfrac{\overline\sigma^2}{2G} \,, \label{SMF}
\end{align}
where the momentum-dependent mass function $M(p^2)$ is determined by the gap equation:
\begin{equation*}
M(p^2) = m_q + r^2(p^2)\,\overline\sigma=m_q + \mathcal{C}(p^2)\,\overline\sigma\ .
\tag{\ref{SMF}a}
\label{M}
\end{equation*}
Here the form factor $\mathcal{C}(p^2)=r^2(p^2)$ is introduced to simplify the expression.

To extend the formulation to finite temperature $T$ and chemical potential $\mu$, the integration over the fourth momentum component is replaced by a Matsubara summation:
\begin{align}
& \int \frac{d^4 p}{(2\pi)^4} F(p_4,\vec p) \quad \to \quad \sint F(p_4,\vec p), \\
& \sint F(p_4,\vec p) \coloneqq
T \sum_{n=-\infty}^{\infty} \int \frac{d^3 p}{(2\pi)^3}
F(\omega_n + i\mu,\vec p)\,,
\end{align}
where $\omega_n = (2n + 1)\pi T$ are the Matsubara frequencies for fermionic modes.

The equilibrium phase corresponds to the stationary point
of the thermodynamic potential as a function of the condensate $\overline\sigma$:
\begin{equation}
\frac{\partial \Omega_{\text{MF}}}{\partial \overline\sigma} = 0\,,
\label{dercondition}
\end{equation}
leading to the gap equation:
\begin{equation}
\overline\sigma = 8 N_c G \sint \frac{M(p^2) \mathcal{C}(p^2)}{p^2 + M^2(p^2)}\,.
\label{gapequation}
\end{equation}

Other relevant physical quantities, such as the chiral condensate $\langle \bar q q \rangle$ for each flavor and the corresponding thermal susceptibilities $\chi$, can be derived from the thermodynamic potential as:
\begin{align}
\langle \bar q q\rangle &= \frac{\partial\Omega}{\partial m_q} = - 4 N_c \sint \frac{M(p^2)}{p^2 + M^2(p^2)}\,, \\
\chi_T &= -\frac{\partial \overline\sigma}{\partial T}\,, \quad
\chi_{\mu} = -\frac{\partial \overline\sigma}{\partial \mu}\,.
\label{qbarq}
\end{align}

Each solution $\overline\sigma_i(T,\mu)$ of Eq.~\eqref{gapequation} depends on the temperature and chemical potential, with $i$ indexing possible multiple solutions. 
The thermal susceptibility $\chi(\overline\sigma_i)$ can be obtained by differentiating both sides of Eq.~\eqref{dercondition} with respect to $T$ (or $\mu$), yielding the following relations:
\begin{align}
\chi_T &= \left( \frac{ \partial^2 \Omega_{\text{MF}} }{\partial T \partial \overline\sigma} \right) \bigg/ \left( \frac{ \partial^2 \Omega_{\text{MF}} }{ \partial \overline\sigma^2 } \right)\,,  \nonumber \\
\chi_{\mu} &= \left( \frac{ \partial^2 \Omega_{\text{MF}} }{ \partial \mu \partial \overline\sigma } \right) \bigg/ \left( \frac{ \partial^2 \Omega_{\text{MF}} }{ \partial \overline\sigma^2 } \right) \,.
\label{chiexp}
\end{align}

\subsection{The Form Factor $\mathcal{C}(p^2)$}

Two commonly used schemes for the form factor $\mathcal{C}(p^2)$ are described here, denoted as FA and FB, respectively. It is well known that the presence of a non-perturbative QCD vacuum, characterized by a nonzero quark condensate $\langle\bar q q\rangle \neq 0$, dynamically generates a momentum-dependent quark mass. This results in a transition from massless current quarks to quasiparticle-like constituent quarks. The spontaneous breaking of chiral symmetry thus leads to a nontrivial dynamical quark mass $M(p^2)$, and the form factor $\mathcal{C}(p^2)$ can be obtained by fitting the lattice QCD (LQCD) results for $M(p^2)$ as shown in \Eqref{M}.

For the FA scheme~\cite{gomezdummChiralQuarkModels2002}, a simple ansatz is adopted:
\begin{equation}
    \mathcal{C}(p^2) = \exp\left(-\frac{p^2}{\Lambda^2}\right)\,.
\label{FA}
\end{equation}
This exponential form provides a good description at low momenta, but it decays too rapidly at high momenta, thereby failing to reproduce the correct ultraviolet behavior.

\begin{figure}
    \centering
    \includegraphics[width=0.8\columnwidth]{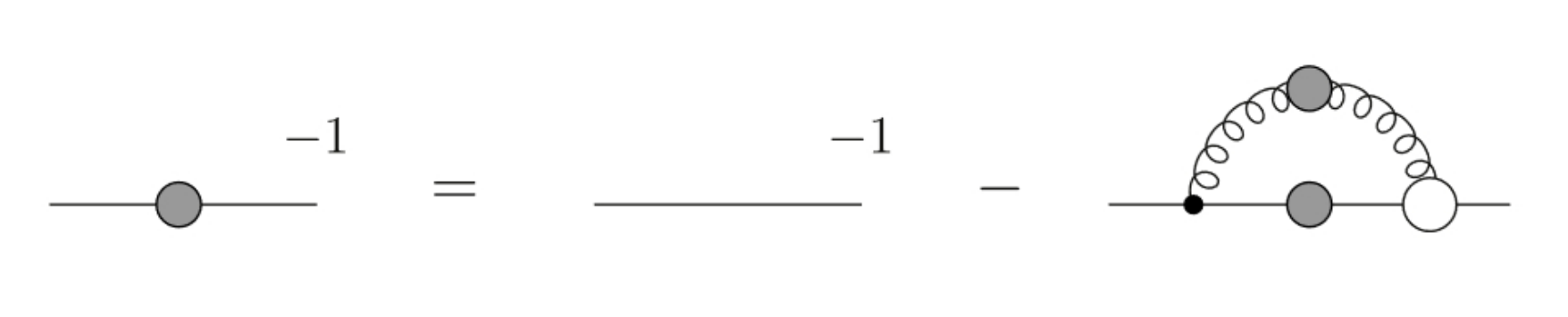}
    \caption{Dyson–Schwinger equation for the quark propagator. The grey blobs with straight lines stand for the full quark propagators, the grey blob with curly line is the gluon propagator, the white blob stands for the full quark-gluon interaction vertex, and the black dots stand for the classical vertex. With the non-local NJL interaction, the form factor $\mathcal{C}(p^2)$ matches the combination of gluon propagator and full quark-gluon vertex, as indicated in the gap equation in Eq.~\eqref{gapequation}.}
    \label{fig-dsequark}
\end{figure}

A refined treatment can be motivated by examining the Dyson–Schwinger equation (DSE) for the quark propagator, as shown in Fig.~\ref{fig-dsequark}. 
The second term on the right-hand side includes the fully dressed quark and gluon propagators, as well as the quark-gluon vertex, which represents the quark self-energy $\Sigma(p)$ in the QCD vacuum. 
As a model for this term, the form factor $\mathcal{C}(p^2)$ should incorporate the chemical potential dependence, i.e., $\mathcal{C}(\omega_n + i\mu, \vec p)$.

At high momentum, a simplification is often taken by using the bare vertex $\gamma_\nu$ (rainbow truncation) and replacing the full gluon propagator by its free form, i.e.,
\begin{equation}
4\pi\alpha_\text{s}(p^2) D_{\mu\nu}(p) \to 4\pi\alpha_\text{s}(p^2) D_{\mu\nu}^\text{free}(p),
\end{equation}
where $\alpha_\text{s}(p^2)$ is the running QCD coupling. In this approximation, the leading nontrivial contribution to the quark self-energy $\Sigma(p^2)$ takes the form:
\begin{equation}\label{quarkselfen}
\Sigma(p^2) = \pi\,\frac{N_\text{c}^2 - 1}{2 N_\text{c}^2 N_\text{f}}\,\frac{\alpha_\text{s}(p^2)}{p^2}\,(3 + \xi)\,\langle\bar\psi\psi\rangle + \delta\Sigma\,,
\end{equation}
where $\xi$ is the gauge parameter and $\delta\Sigma$ denotes the subleading corrections.

Based on this analysis, the FB scheme for the form factor is proposed as\cite{hellDynamicsThermodynamicsNonlocal2009}:
\begin{equation}
\mathcal{C}(p^2) = 
\begin{cases}
\exp\left(-\dfrac{p^2 d^2}{2}\right), & \text{for } p^2 < \varGamma^2, \\
\text{const.} \cdot \dfrac{\alpha_\text{s}(p^2)}{p^2}, & \text{for } p^2 \geq \varGamma^2.
\end{cases}
\end{equation}

In this work, we adopt the simpler FA form given in Eq.~\eqref{FA}. The model parameters to be determined are the coupling strength $G$ in Eq.~\eqref{Sint}, the current quark mass $m_q$, and the cutoff parameter $\Lambda$ in Eq.~\eqref{FA}. These parameters can be fixed by fitting to physical observables such as the dynamical quark mass $M(p^2)$, the empirical value of the pion decay constant, and the pion mass $m_\pi$. 
Following Refs.~\cite{dummStronginteractionMatterExtreme2021,dummCovariantNonlocalChiral2006}, we choose the parameter set:
\begin{equation}
G\Lambda^2 = 20.65\,, \quad m_q = 5.74~\text{MeV}\,, \quad \Lambda = 752~\text{MeV}\,.
\end{equation}

\section{Calculation and Numerical Results}\label{sec:phasediagram}
\subsection{Real Chemical Potential}

As discussed earlier, the quark condensate $\langle\bar q q\rangle$ is dynamically generated at low temperature $T$ and low chemical potential $\mu$. In this regime, quarks behave as quasiparticles with a momentum-dependent mass $M(p^2)$—a consequence of dynamical chiral symmetry breaking. As $T$ or $\mu$ increases, the condensate melts, and the dynamical mass vanishes. 
From Eq.~\eqref{M}, we observe that the solution $\overline\sigma$ of the gap equation differs from $M(p^2=0)$ only by the explicit chiral symmetry breaking term due to the current quark mass $m_q$. Therefore, $\overline\sigma$ can also be used as an order parameter for the chiral phase transition.
Moreover, $\overline\sigma$ corresponds to the stationary point of the thermodynamic potential $\Omega$, making it instructive to examine how $\Omega$ evolves with temperature and chemical potential.

\begin{figure}
    \centering
    \includegraphics[width=0.8\linewidth]{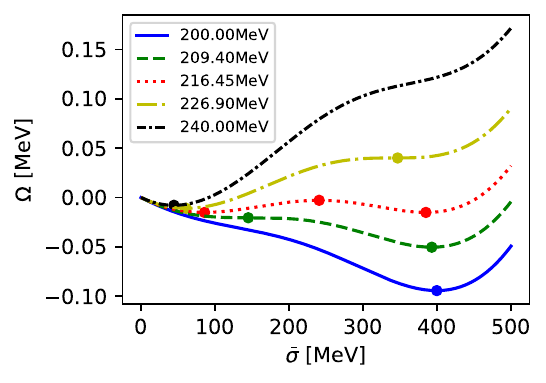}
    \caption{Thermodynamic potential $\Omega$ as a function of $\overline\sigma$ at $T = 80$~MeV for several values of chemical potential $\mu$. Solutions of the gap equation are indicated by dots.}
    \label{fig:omegat80}
\end{figure}

The obtained result is illustrated in Fig.~\ref{fig:omegat80}, for the case of $T = 80$~MeV and at different chemical potentials.
It shows that, at $\mu = 200$~MeV, the thermodynamic potential $\Omega$ exhibits a single global minimum at $\overline\sigma = 399.7$~MeV, referred to as the \textbf{positive Nambu solution}.
As $\mu$ increases to $209.4$~MeV, a new pair of extrema emerges: a local maximum and a local minimum. The newly formed minimum is identified as the \textbf{Wigner solution}.
Upon further increase in $\mu$, the Wigner solution becomes the global minimum, and the system undergoes a first-order phase transition at $\mu = 216.45$~MeV, where $\overline\sigma$ exhibits a discontinuous jump.
Eventually, the positive Nambu solution vanishes at $\mu = 226.9$~MeV. 

In contrast, at higher temperatures (e.g., $T = 100$~MeV), $\overline\sigma$ varies smoothly with $\mu$, indicating a crossover rather than a genuine phase transition. 
Thus, a \textbf{critical endpoint (CEP)} exists at which the first-order phase transition line terminates. 
At the CEP, the three extrema of $\Omega$—the positive Nambu solution, the Wigner solution, and the intermediate local maximum—coalesce into a single degenerate solution.

\begin{figure}[htbp]
    \centering
    \begin{subfigure}[b]{\linewidth}
        \centering
        \includegraphics[width=0.7\linewidth]{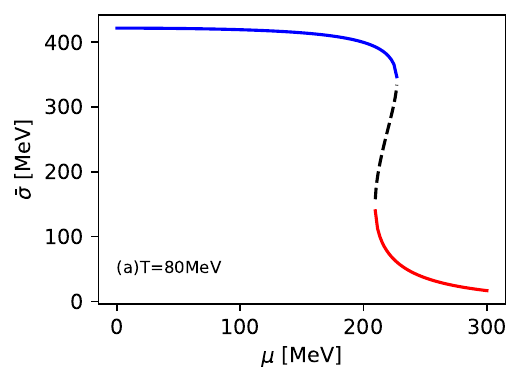}
        \captionsetup{margin=1.5cm}
    \end{subfigure}
    \hfill~\\
    \begin{subfigure}[b]{\linewidth}
        \centering
        \includegraphics[width=0.7\linewidth]{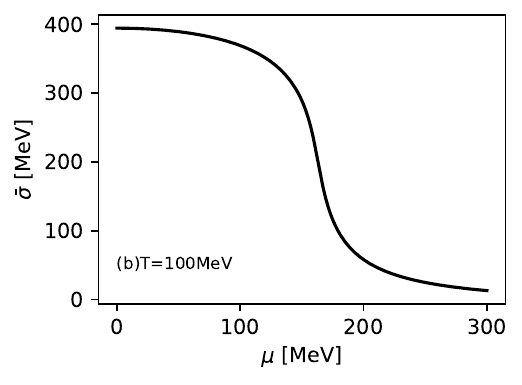}
        \captionsetup{margin=1.5cm}
    \end{subfigure}
    \caption{Solutions of the gap equation $\overline\sigma(\mu)$ at two different temperatures.}
    \label{fig:sigma}
\end{figure}

Fig.~\ref{fig:sigma} shows the solutions of the gap equation as a function of $\mu$ in cases of $T = 80$~MeV and $T = 100$~MeV.

\begin{figure}[htbp]
    \centering
    \includegraphics[width=0.7\linewidth]{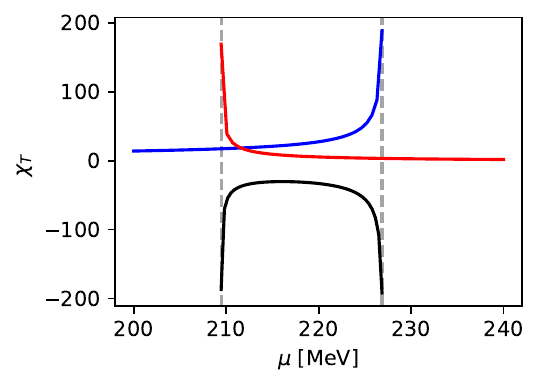}
    \caption{Thermal susceptibility $\chi_T$ as a function of the chemical potential $\mu$ at $T = 80$~MeV. The blue, red, and black curves represent $\chi_T$ of the positive Nambu solution, the Wigner solution, and the local maximum solution, respectively.}
    \label{fig:chimui0}
\end{figure}

Solutions to the gap equation typically emerge or vanish in pairs, consisting of a local maximum \big($\frac{\partial^2 \Omega}{\partial \overline\sigma^2} < 0$\big) and a local minimum \big($\frac{\partial^2 \Omega}{\partial \overline\sigma^2} > 0$\big). When such a pair coalesces, the second derivative vanishes \big($\frac{\partial^2 \Omega}{\partial \overline\sigma^2} = 0$\big) , leading to a divergence in the susceptibility $\chi_T$ [see Eq.~\eqref{chiexp}]. 
As shown in Fig.~\ref{fig:chimui0}, divergences at $\mu = 209.4$~MeV and $\mu = 226.9$~MeV signal the presence of \textbf{spinodal points} in the phase diagram. 
These points delineate the \textbf{spinodal lines}, which enclose the \textbf{spinodal region}—a metastable domain associated with supercooling and superheating phenomena~\cite{kartheinDescriptionFirstOrder2024,PhysRevD.89.074041,PhysRevD.94.094030}.

For physical quark masses and low chemical potentials ($\mu < \mu_\text{CEP}$), the chiral transition is a crossover, characterized by a smooth variation of the order parameter. The chiral crossover temperature $T_c(\mu_B)$ can be extracted from the temperature dependence of $\chi_T$, where its peak signals $T_c$ at a given $\mu_B$. 

\begin{figure}
    \centering
    \includegraphics[width=0.8\linewidth]{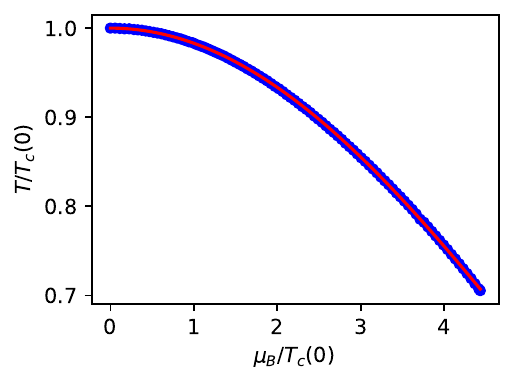}
    \caption{Chiral crossover line extracted from the thermal susceptibility $\chi_T$ in the nonlocal NJL model, which is rescaled by the pseudo-critical temperature $T_c(0) = 127.7~\text{MeV}$ at vanishing $\mu$.}
    \label{fig:crossoverfit}
\end{figure}

The curvature of the crossover line in the $(T, \mu_B)$ plane is a key observable, which can be a benchmark for the model:
\begin{equation}
    \frac{T_c(\mu_B)}{T_c(0)} = 1 - \kappa \left(\frac{\mu_B}{T_c(0)}\right)^2 + \lambda \left(\frac{\mu_B}{T_c(0)}\right)^4 + \dots
\end{equation}
% where $T_c(0)=127.7~\text{MeV}$.
Fig.~\ref{fig:crossoverfit} shows the fitting result for $T_c(\mu_B)$ obtained from the maxima of $\chi_T$ in the range $\mu_B = 0$–566.2~MeV. The fitted curvature is $\kappa = 0.01708(2)$, which is consistent with the predictions from previous studies in 2 flavor: $\kappa = 0.0175(7)$~\cite{Gao:2020fbl}, $\kappa = 0.0160(1)$~\cite{Pawlowski_2014}. Additional lattice QCD results for $\kappa$ can be found in Ref.~\cite{D_Elia_2019}.

%result from 2-flavor DSE studies, $\kappa = 0.0175(7)$~\cite{Gao:2020fbl}. 

\begin{figure}
    \centering
    \includegraphics[width=0.8\linewidth]{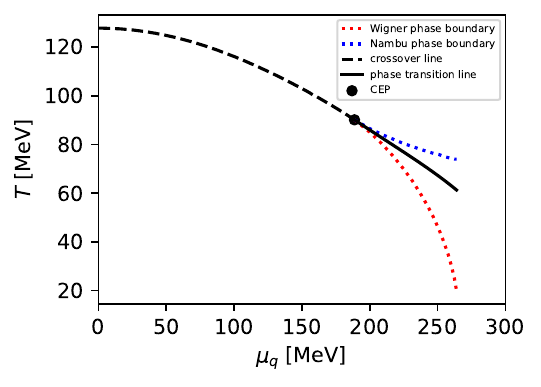}
    \caption{The obtained phase diagram in the $(T, \mu)$ plane in nonlocal NJL model. The red and blue dotted lines (the spinodal lines) denote the boundaries of the Wigner phase and the positive Nambu phase, respectively, where each phase merges with the metastable phase, as shown in Figure~\ref{fig:omegat80}.}
    \label{fig:phasediagram}
\end{figure}

The obtained phase diagram is shown in Fig.~\ref{fig:phasediagram}. It is clear that the critical end point is located at the intersection of spinodal lines at
% \begin{align}
% & T_\text{CEP} = 90.10~\text{MeV}, \nonumber \\
% & \mu_\text{CEP} = 188.73~\text{MeV}.
% \end{align}
\begin{equation}
 T_\text{CEP} = 90.10~\text{MeV}, \quad
 \mu_\text{q,CEP} = 188.73~\text{MeV}.
\end{equation}

\subsection{Complex Chemical Potential}

Before studying the nonlocal NJL model with a complex chemical potential, let us briefly review the concept of Lee--Yang zeros. In a grand canonical ensemble, the partition function can be expressed as a polynomial in the fugacity $\zeta = e^{\mu/T}$. Therefore, the distribution of the zeros of the fugacity encodes all thermodynamic information of the system. The zeroes coalesce into branch cuts emanating from the so called Lee--Yang edge singularities(LYEs). A second-order (first-order) phase transition occurs when the real axis of $\mu$ (or $T$) crosses the Lee--Yang edge singularity(branch cuts).

When the chemical potential is analytically continued into the complex plane, the thermodynamic potential $\Omega(T,\mu,\overline\sigma)$ also becomes complex. Consequently, the solutions of the gap equation are distributed in the complex plane, and the concept of a minimum becomes ill-defined, since complex values cannot be ordered.

To locate second-order phase transitions, i.e., the LYES, several representative methods have been employed in recent studies\cite{basarUniversalityLeeYangSingularities2021,wanLeeYangEdgeSingularities2024,deforcrandQCDPhaseDiagram2002a,An:2016lni}. One of them involves analyzing the radius of the convergence of the Taylor expansion of the thermodynamic potential $\Omega$ around $\mu=0$ up to order $N$:
\begin{equation}
    \Omega^E_N(T,\mu) = \sum_{n=1}^N \frac{1}{n!} \left. \frac{\partial^n \Omega}{\partial \mu^n} \right|_{\mu=0} \mu^n.
\end{equation}
Alternatively, a resummation scheme for the thermodynamic potential, denoted by $\Omega^R_N(T,\mu)$, provides more direct insight into the location of LYES. Because even at finite order, $\Omega^R_N$ includes infinite powers of $\mu$~\cite{mondalLatticeQCDEquation2022}.

In this article, we propose an improved method for locating the LYES. As mentioned earlier, the CEP lies at the intersection of spinodal lines, where three solutions of the gap equation coalesce. When $\mu_I = 0$, the spinodal lines can be identified by counting the number of real-valued solutions to the gap equation. However, when analytically continued into the complex $\overline\sigma$ plane, the total number of solutions remains constant—these solutions merely depart from or approach the real axis. The situation becomes more complicated when $\mu_I \neq 0$, as all solutions may become complex. Consequently, a modified approach is required to identify the LYES in the complex chemical potential plane.

A key observation is that the spinodal line separates regions in the phase diagram, which are characterized by different shapes of the thermodynamic potential surface. 

When extended into the complex $\overline\sigma$ plane, the thermodynamic potential $\Omega(\overline\sigma; T, \mu)$ becomes a complex-valued function defined over a two-dimensional domain. It can be expressed as $\Omega = \Omega_R + i\Omega_I$, where both $\Omega_R$ and $\Omega_I$ are real functions of $\overline\sigma$, $T$, and $\mu$. Assuming that $\Omega$ is analytic in $\overline\sigma$, the real part $\Omega_R(\overline\sigma)\vert_{T, \mu}$ alone suffices to determine the full structure of the potential surface, due to the Cauchy–Riemann conditions.

\begin{figure}
    \centering
    \includegraphics[width=0.8\columnwidth]{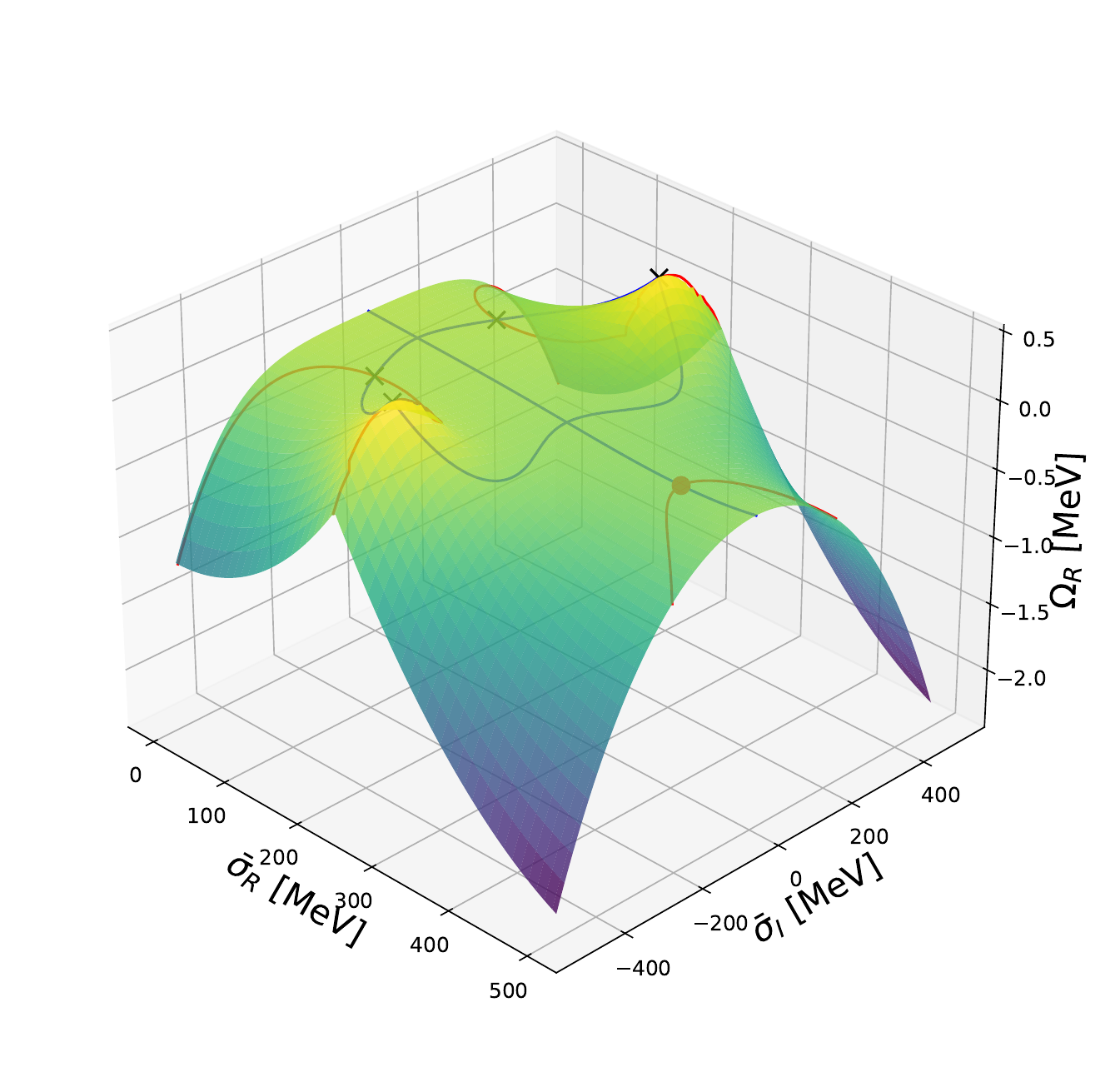}
    \caption{Thermodynamic potential $\Omega$ in the complex plane of $\overline\sigma$ at T = 80 MeV and $\mu=180$ MeV. The red solid line corresponds to solutions of $\textbf{Re}(\frac{\partial \Omega}{\partial \overline\sigma})=0$, the blue line corresponds to solutions of $\textbf{Im}(\frac{\partial \Omega}{\partial \overline\sigma})=0$. The red point is the physical stationary point(the positive Nambu solution).}
    \label{omegat80mu200c}
\end{figure}

Figure~\ref{omegat80mu200c} displays the real part of the thermodynamic potential, $\Omega_R(\overline\sigma)\vert_{T, \mu}$, in the complex $\overline\sigma$ plane at $T = 80~\text{MeV}$, $\mu_R = 200~\text{MeV}$, and $\mu_I/\pi T = 0$. 
The condition for stationary points, originally given by \Eqref{dercondition}, becomes:
\begin{align}
    \frac{\partial \Omega}{\partial \overline\sigma} 
    &= \frac{\partial \Omega}{\partial \overline\sigma_R} 
    = \frac{\partial \Omega_R}{\partial \overline\sigma_R} + i \frac{\partial \Omega_I}{\partial \overline\sigma_R} 
    = \frac{\partial \Omega_R}{\partial \overline\sigma_R} - i \frac{\partial \Omega_R}{\partial \overline\sigma_I} = 0, \nonumber \\
    &\Rightarrow \quad \frac{\partial \Omega_R}{\partial \overline\sigma_R} = 0 ,\quad \frac{\partial \Omega_R}{\partial \overline\sigma_I} = 0.
\end{align}
The red and blue lines in the figure represent the solutions to the real and imaginary parts of the gap equation, corresponding to the conditions $\partial \Omega_R/\partial \overline\sigma_R = 0$ and $\partial \Omega_R/\partial \overline\sigma_I = 0$, respectively. These curves jointly determine the stationary points and reflect the underlying geometry of the potential surface.

The red point marks the physical stationary point on the real axis, while the black forks indicate other stationary solutions in the complex $\overline\sigma$ plane. 
For $\mu_I = 0$, the condition for a physical solution is that $\frac{\partial^2 \Omega_R}{\partial \overline\sigma_R^2} > 0$, which further requires $\frac{\partial^2 \Omega_R}{\partial \overline\sigma_I^2} < 0$. This implies that the physical solution now appears as a saddle point on the surface of $\Omega_R$.

\begin{figure}[htbp]
    \centering
    \captionsetup[subfigure]{skip=0pt,justification=centering, font=small}
    \begin{subfigure}[b]{0.45\columnwidth}
        \includegraphics[width=\textwidth]{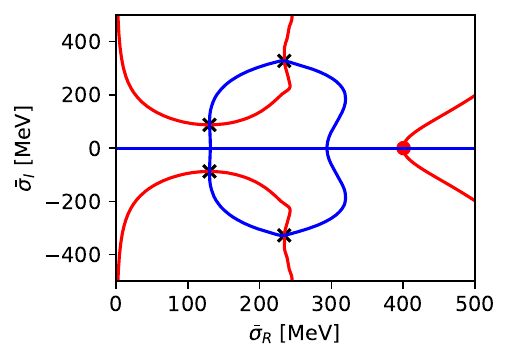} 
        \caption{(80.00,200.00)MeV}
        \label{realgapeqcross-a}
    \end{subfigure}
    \hfill~\hspace{-0.05\columnwidth}
    \begin{subfigure}[b]{0.45\columnwidth}
        \includegraphics[width=\textwidth]{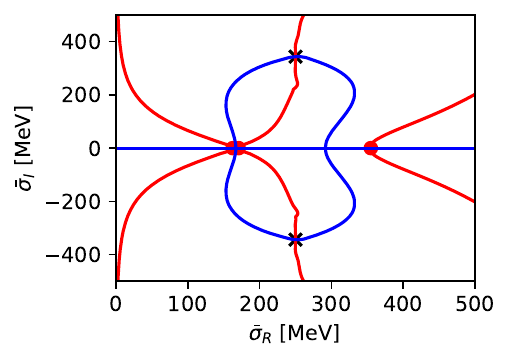} 
        \caption{(85.03,200.00)MeV}
        \label{realgapeqcross-b}
    \end{subfigure}
    \vspace{1.5em}
    
    \begin{subfigure}[b]{0.45\columnwidth}
        \includegraphics[width=\textwidth]{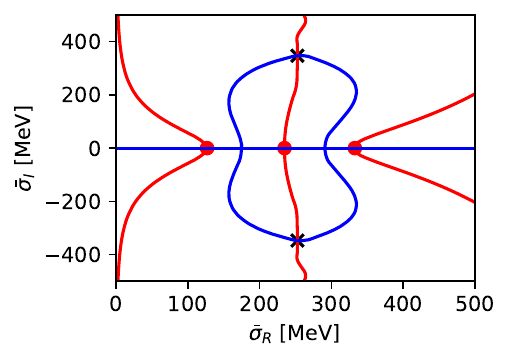} 
        \caption{(86.00,200.00)MeV}
        \label{realgapeqcross-c}
    \end{subfigure}
    \hfill~\hspace{-0.05\columnwidth}
    \begin{subfigure}[b]{0.45\columnwidth}
        \includegraphics[width=\textwidth]{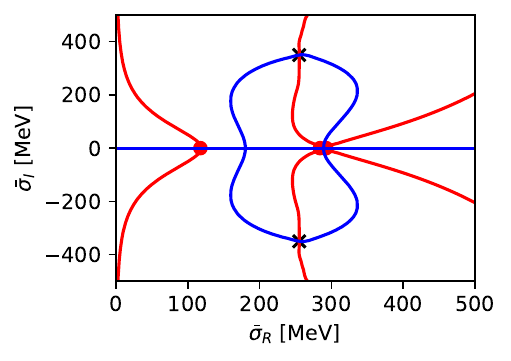} 
        \caption{(86.59,200.00)MeV}
        \label{realgapeqcross-d}
    \end{subfigure}
    \vspace{1.5em}
    
    \begin{subfigure}[b]{0.45\columnwidth}
        \includegraphics[width=\textwidth]{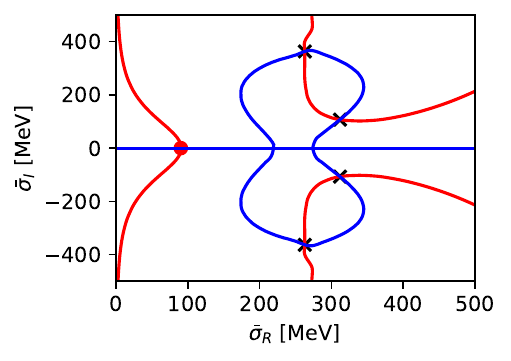} 
        \caption{(90.00,200.00)MeV}
        \label{realgapeqcross-e}
    \end{subfigure}
    \hfill~\hspace{-0.05\columnwidth}
    \begin{subfigure}[b]{0.45\columnwidth}
        \includegraphics[width=\textwidth]{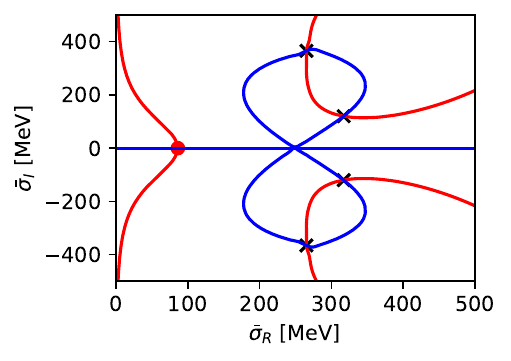} 
        \caption{(90.94,200.00)MeV}
        \label{realgapeqcross-f}
    \end{subfigure}
    \vspace{1.5em}
    
    \begin{subfigure}[b]{0.45\columnwidth}
        \includegraphics[width=\textwidth]{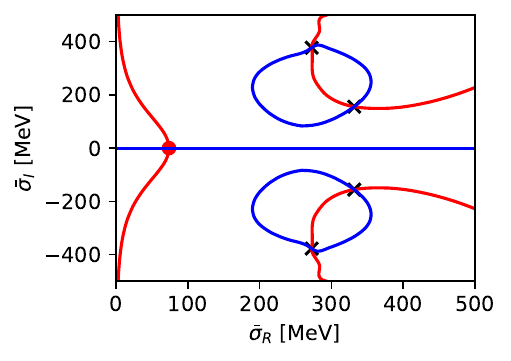} 
        \caption{(94.00,200.00)MeV}
        \label{realgapeqcross-g}
    \end{subfigure}
    \hfill~\hspace{-0.05\columnwidth}
    \begin{subfigure}[b]{0.45\columnwidth}
        \includegraphics[width=\textwidth]{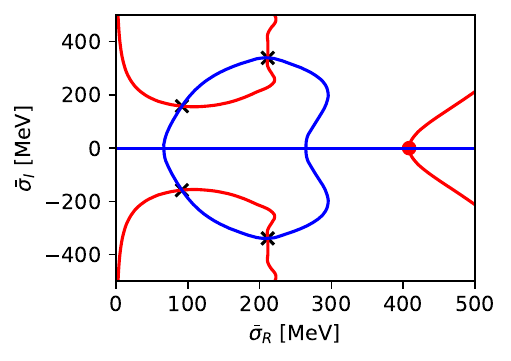} 
        \caption{(80.00,180.00)MeV}
        \label{realgapeqcross-h}
    \end{subfigure}
    \vspace{1.5em}

    \begin{subfigure}[b]{0.45\columnwidth}
        \includegraphics[width=\textwidth]{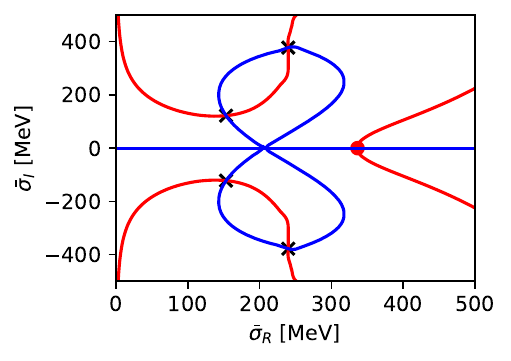} 
        \caption{(90.15,180.00)MeV}
        \label{realgapeqcross-i}
    \end{subfigure}
    \hfill~\hspace{-0.05\columnwidth}
    \begin{subfigure}[b]{0.45\columnwidth}
        \includegraphics[width=\textwidth]{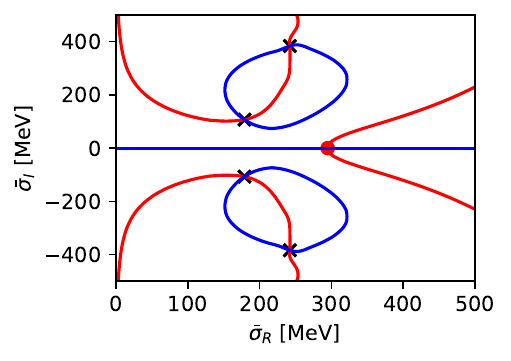} 
        \caption{(92.00,180.00)MeV}
        \label{realgapeqcross-j}
    \end{subfigure}
    \vspace{1.5em}

    \begin{subfigure}[b]{0.45\columnwidth}
        \includegraphics[width=\textwidth]{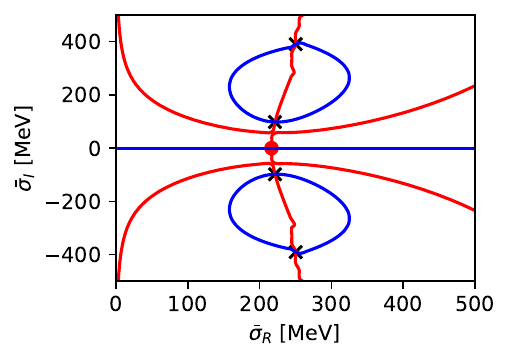} 
        \caption{(93.40,180.00)MeV}
        \label{realgapeqcross-k}
    \end{subfigure}
    \hfill~\hspace{-0.05\columnwidth}
    \begin{subfigure}[b]{0.45\columnwidth}
        \includegraphics[width=\textwidth]{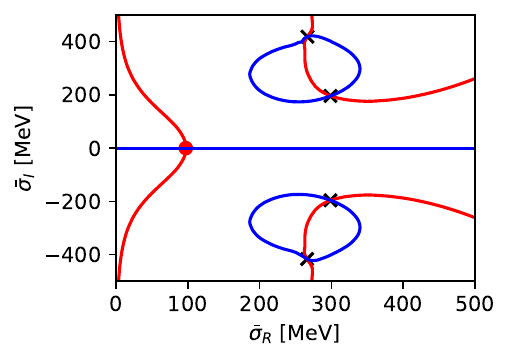} 
        \caption{(100.00,180.00)MeV}
        \label{realgapeqcross-l}
    \end{subfigure}

    \caption{Solutions of the gap equation in the complex $\overline\sigma$ plane at $\mu_I /\pi T= 0$. Red lines correspond to the real part of the gap equation, while blue lines correspond to the imaginary part.  }
    \label{realgapeqcross}
\end{figure}

For $\mu_I/\pi T = 0$, the potential shapes characterized by the solutions of the real and imaginary parts of the gap equation are shown in Figure~\ref{realgapeqcross}. 
Figures~\ref{realgapeqcross-a} and \ref{realgapeqcross-h} represent the typical potential shapes at some low temperature and chemical potential, while Figures~\ref{realgapeqcross-g} and \ref{realgapeqcross-l} illustrate those at high temperature and chemical potential.

Fixing $\mu_R = 200~\text{MeV}$, Figures~\ref{realgapeqcross-a} through \ref{realgapeqcross-g} display the evolution of the potential shape across a first-order phase transition as the temperature increases from $80~\text{MeV}$ to $100~\text{MeV}$. 
In contrast, fixing $\mu_R = 180~\text{MeV}$, Figures~\ref{realgapeqcross-h} through \ref{realgapeqcross-l} demonstrate the shape evolution in a crossover region over the same temperature range.

\begin{figure}
    \centering
    \captionsetup[subfigure]{skip=0pt,justification=centering, font=small}
    \begin{subfigure}[b]{0.45\columnwidth}
        \includegraphics[width=\textwidth]{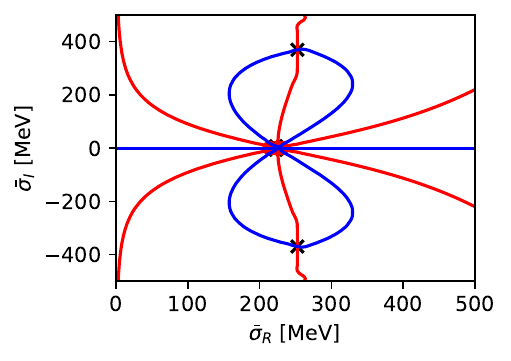} 
        \caption{CEP}
        \label{realgapeqcross-CEP}
    \end{subfigure}
    \hfill
    \begin{subfigure}[b]{0.45\columnwidth}
        \includegraphics[width=\textwidth]{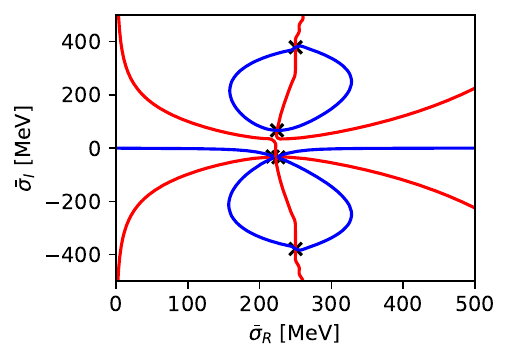} 
        \caption{LYEs}
        \label{imaggapeqcross-LYEs}
    \end{subfigure}
    \caption{Solutions of the gap equation in the complex $\overline\sigma$ plane at CEP and LYEs. \\(a)CEP:$T = 90.10~\text{MeV}$, $\mu_{R} = 188.73~\text{MeV}$, $\mu_{I}/\pi T = 0$. (b)LYEs:$T = 91.29~\text{MeV}$, $\mu_{R} = 185.57~\text{MeV}$, $\mu_{I}/\pi T = 0.001$.}
    \label{gapeqcrossCEPandLYEs}
\end{figure}

Figures~\ref{realgapeqcross-b}, \ref{realgapeqcross-d}, \ref{realgapeqcross-f}, \ref{realgapeqcross-i}, and \ref{realgapeqcross-k} correspond to the critical points where the potential shape qualitatively changes. Lines formed by connecting these points in the phase diagram are referred to as \emph{shape-shifting lines}, which include the spinodal lines. All such lines intersect at the critical end point (CEP) or the Lee--Yang edge singularities (LYEs), leading to a distinct potential structure at those points, as shown in Figure~\ref{gapeqcrossCEPandLYEs}.

This categorization allows one to determine the location of $(T, \mu)$ in the phase diagram based on the shape of the thermodynamic potential. For instance, the configurations shown in Figures~\ref{realgapeqcross-c} and \ref{realgapeqcross-e} indicate that $(T, \mu)$ lies near the first-order phase transition line with $\mu > \mu_\text{CEP}$, while Figure~\ref{realgapeqcross-j} suggests that $(T, \mu)$ is in the crossover region with $\mu < \mu_\text{CEP}$.

\begin{figure}[htbp]
    \centering
    \captionsetup[subfigure]{skip=0pt,justification=centering, font=small}
    \begin{subfigure}[b]{0.45\columnwidth}
        \includegraphics[width=\textwidth]{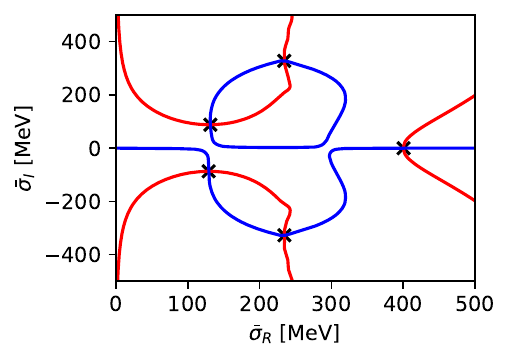} 
        \caption{(80.00,200.00)MeV}
        \label{imaggapeqcross-a}
    \end{subfigure}
    \hfill~\hspace{-0.05\columnwidth}
    \begin{subfigure}[b]{0.45\columnwidth}
        \includegraphics[width=\textwidth]{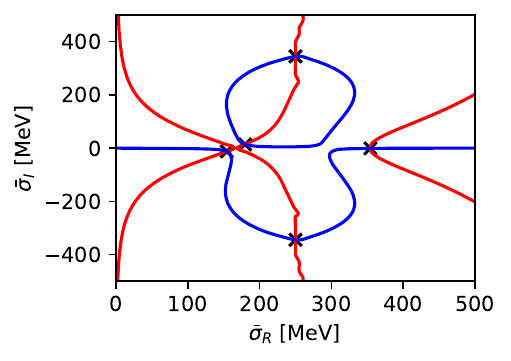} 
        \caption{(85.03,200.00)MeV}
        \label{imaggapeqcross-b}
    \end{subfigure}
    \vspace{1.5em}

    \begin{subfigure}[b]{0.45\columnwidth}
        \includegraphics[width=\textwidth]{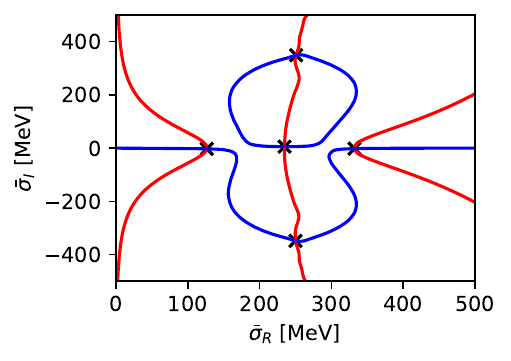} 
        \caption{(86.00,200.00)MeV}
        \label{imaggapeqcross-c}
    \end{subfigure}
    \hfill~\hspace{-0.05\columnwidth}
    \begin{subfigure}[b]{0.45\columnwidth}
        \includegraphics[width=\textwidth]{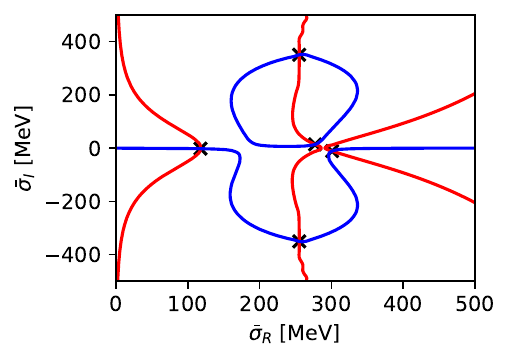} 
        \caption{(86.59,200.00)MeV}
        \label{imaggapeqcross-d}
    \end{subfigure}
    \vspace{1.5em}
    
    \begin{subfigure}[b]{0.45\columnwidth}
        \includegraphics[width=\textwidth]{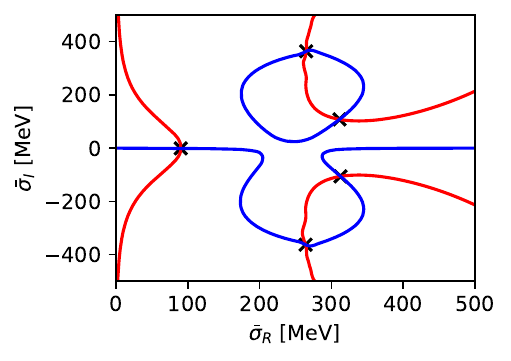} 
        \caption{(90.00,200.00)MeV}
        \label{imaggapeqcross-e}
    \end{subfigure}
    \hfill~\hspace{-0.05\columnwidth}
    \begin{subfigure}[b]{0.45\columnwidth}
        \includegraphics[width=\textwidth]{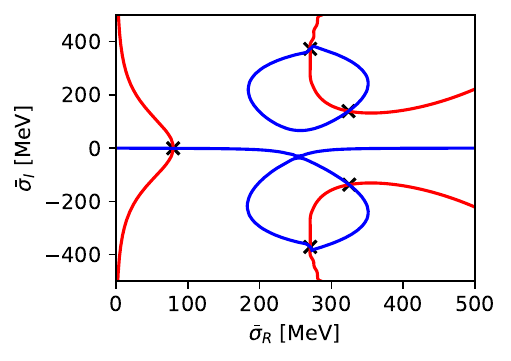} 
        \caption{(92.40,200.00)MeV}
        \label{imaggapeqcross-f}
    \end{subfigure}
    \vspace{1.5em}
    
    \begin{subfigure}[b]{0.45\columnwidth}
        \includegraphics[width=\textwidth]{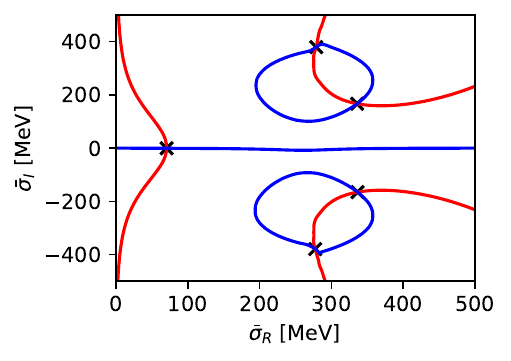} 
        \caption{(95.00,200.00)MeV}
        \label{imaggapeqcross-g}
    \end{subfigure}
    \hfill~\hspace{-0.05\columnwidth}
    \begin{subfigure}[b]{0.45\columnwidth}
        \includegraphics[width=\textwidth]{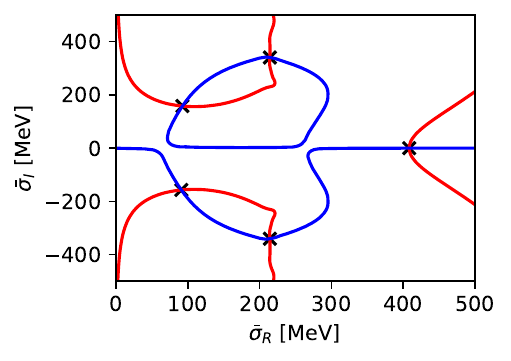} 
        \caption{(80.00,180.00)MeV}
        \label{imaggapeqcross-h}
    \end{subfigure}
    \vspace{1.5em}

    \begin{subfigure}[b]{0.45\columnwidth}
        \includegraphics[width=\textwidth]{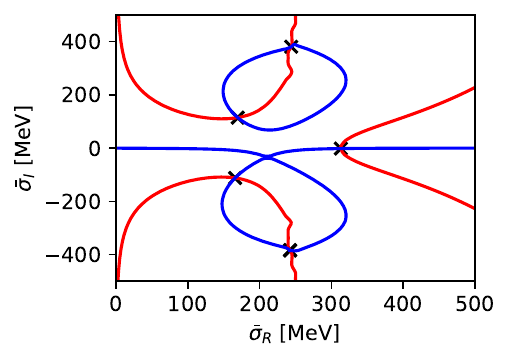} 
        \caption{(91.32,180.00)MeV}
        \label{imaggapeqcross-i}
    \end{subfigure}
    \hfill~\hspace{-0.05\columnwidth}
    \begin{subfigure}[b]{0.45\columnwidth}
        \includegraphics[width=\textwidth]{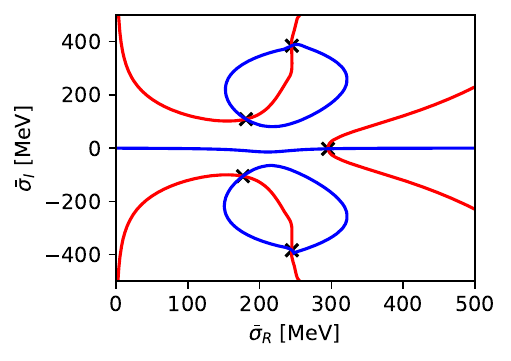} 
        \caption{(92.00,180.00)MeV}
        \label{imaggapeqcross-j}
    \end{subfigure}
    \vspace{1.5em}

    \begin{subfigure}[b]{0.45\columnwidth}
        \includegraphics[width=\textwidth]{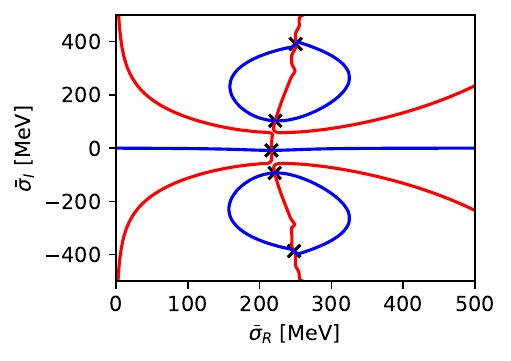} 
        \caption{(93.40,180.00)MeV}
        \label{imaggapeqcross-k}
    \end{subfigure}
    \hfill~\hspace{-0.05\columnwidth}
    \begin{subfigure}[b]{0.45\columnwidth}
        \includegraphics[width=\textwidth]{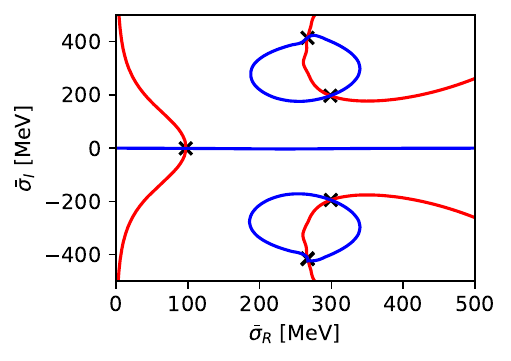} 
        \caption{(100.00,180.00)MeV}
        \label{imaggapeqcross-l}
    \end{subfigure}
    
    \caption{Solutions of the gap equation in the complex $\overline\sigma$ plane at $\mu_I /\pi T= 0.001$. Red lines correspond to the real part of the gap equation, while blue lines correspond to the imaginary part.}
    \label{imaggapeqcross}
\end{figure}

Figure~\ref{imaggapeqcross} presents a similar classification of the potential shapes in the case $\mu_I/\pi T = 0.001$.


\begin{figure}[htbp]
    \begin{subfigure}[b]{0.4\textwidth}
        \includegraphics[width=0.9\textwidth]{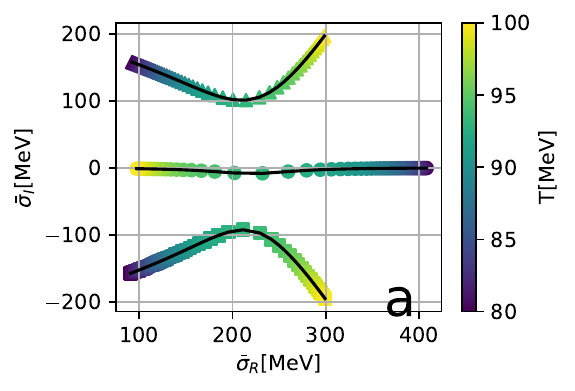} 
        \label{solutiontrajectory-a}
    \end{subfigure}
    \hfill
    \begin{subfigure}[b]{0.4\textwidth}
        \includegraphics[width=0.9\textwidth]{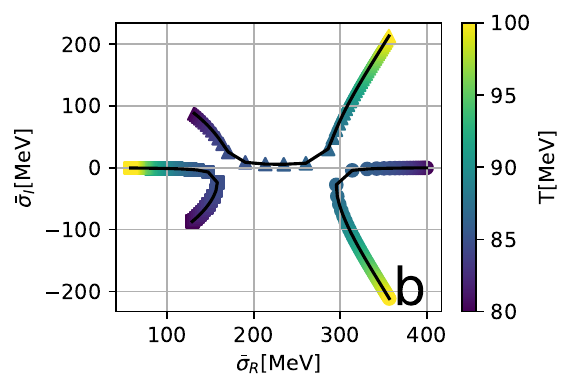} 
        \label{solutiontrajectory-b}
    \end{subfigure}
    \hfill
    \begin{subfigure}[b]{0.4\textwidth}
        \includegraphics[width=0.9\textwidth]{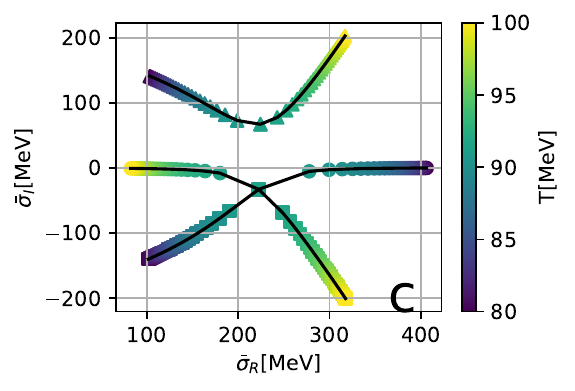} 
        \label{solutiontrajectory-c}
    \end{subfigure}
    \caption{Trajectories of the gap equation solutions at $\mu_I/\pi T = 0.001$. The temperature increases from 80 to 100 MeV, represented by a color gradient. (a) $\mu_R = 180$ MeV, corresponding to a crossover. (b) $\mu_R = 200$ MeV, where a first-order transition occurs.(c) $\mu_R = 185.57$ MeV, where the trajectory passes through the LYEs. Each solution branch is depicted with a distinct marker shape.}
    \label{solutiontrajectory}
\end{figure}

When $T$ or $\mu$ varies, the solutions of the gap equation trace out trajectories in the complex $\overline\sigma$ plane. As the shape of the thermodynamic potential $\Omega$ evolves distinctly across first-order phase transitions compared to crossovers, the trajectories of the solutions exhibit distinct patterns. Representative examples at $\mu_I/\pi T = 0.001$ are shown in Figure~\ref{solutiontrajectory}.

In Fig.\ref{solutiontrajectory}\textbf{a}, the temperature varies from 80 to 100 MeV with a fixed $\mu_R = 180$ MeV, corresponding to the crossover region. In contrast, Fig.\ref{solutiontrajectory}\textbf{b} depicts the case where $\mu_R = 200$ MeV, indicating a first-order phase transition. Keeping $\mu_I/\pi T = 0.001$ and $T \in [80,100]$ MeV, the trajectories of the Nambu and Wigner solutions intersect at $\mu_R = 185.57~\text{MeV}$ as shown in Fig.\ref{solutiontrajectory}\textbf{c}. When $\mu_R$ changes, the solution trajectories of the Nambu and Wigner solutions bifurcate.

Figure~\ref{gapeqcrossCEPandLYEs} and Figure~\ref{solutiontrajectory} imply that the positive Nambu solution $\overline\sigma_1$ coalesces with the Wigner solution $\overline\sigma_2$ at the CEP or more generally at the LYEs. The coalescence of these solutions implies that
\begin{equation}
\left. \frac{\partial^2 \Omega}{\partial \overline\sigma^2} \right|_{(\overline\sigma_l, T_l, \mu_l)} = 0,
\label{LYEsder}
\end{equation}
where $(T_l, \mu_l, \overline\sigma_l)$ are the temperature, chemical potential, and coalesced order parameter at the LYEs(including CEP), respectively.
This is guaranteed by the implicit function theorem. Below we present a general proof, allowing for complex values of $T$ as well.

We begin by introducing the notation:
\begin{equation}
    F(\overline\sigma; T, \mu) := \frac{\partial \Omega(\overline\sigma; T, \mu)}{\partial \overline\sigma}.
\end{equation}
From physical considerations, the thermodynamic potential $\Omega$ is expected to be continuously differentiable with respect to $(\overline\sigma, T, \mu)$. Upon complexification of the parameters, it is further assumed to be holomorphic. Consequently, $F(\overline\sigma; T, \mu)$ is also holomorphic in all its arguments.
It is important to note that the thermodynamic potential evaluated at stationary points (including metastable ones), defined as $\Omega(T, \mu):=\Omega(\overline\sigma; T, \mu)\big|_{F(\overline\sigma; T, \mu) = 0}$, may exhibit non-analytic behavior at spinodal lines. However, this is not the function under consideration in the present analysis.

Since we are interested only in stationary points of the potential, we consider solutions of the equation
\begin{equation}
    F(\overline\sigma; T, \mu) = 0,
\end{equation}
as implied by \Eqref{dercondition}.

To prove \Eqref{LYEsder} at the Lee--Yang edge singularities (LYEs), we proceed by contradiction. Suppose, contrary to the claim, that
\begin{equation}
    \left. \frac{\partial F}{\partial \overline\sigma} \right|_{(\overline\sigma_l, T_l, \mu_l)} \neq 0.\label{eq:nondegenerate-assumption}
\end{equation}

% \cYL{Put point 1 and 2 as condition (also use more physical terms), then clarify that ``we want to prove point 3 does not hold at LYE.''}

We define an auxiliary function $H: \mathbb{C} \times \mathbb{C}^2 \to \mathbb{C} \times \mathbb{C}^2$ as
\begin{equation}
    H(\overline\sigma; T, \mu) := \big( F(\overline\sigma; T, \mu),\; T,\; \mu \big).
\end{equation}

The Jacobian matrix of $H$ evaluated at the point $(\overline\sigma_l, T_l, \mu_l)$ is given by
\begin{align}
    J_H = \begin{pmatrix}
             \dfrac{\partial F}{\partial \overline\sigma} & \dfrac{\partial F}{\partial T} & \dfrac{\partial F}{\partial \mu} \\
             0 & 1 & 0 \\
             0 & 0 & 1
          \end{pmatrix}, \\
    \left. \det(J_H) \right|_{(\overline\sigma_l, T_l, \mu_l)} = \left. \frac{\partial F}{\partial \overline\sigma} \right|_{(\overline\sigma_l, T_l, \mu_l)} \neq 0.
\end{align}

By the Complex Inverse Function Theorem, $H$ is locally biholomorphic near $(\overline\sigma_l, T_l, \mu_l)$. Hence, in a neighborhood $U$, there exists a unique inverse function
\begin{equation}
    H^{-1}(F(\overline\sigma; T, \mu), T, \mu) = (\overline\sigma, T, \mu).
    \label{Hinverse1}
\end{equation}

Given that \(F(\overline\sigma; T, \mu) = 0\), we obtain
\begin{equation}
    H^{-1}(0,\; T,\; \mu) = (\overline\sigma,\; T,\; \mu),
    \label{Hinverse2}
\end{equation}
which implies the existence of a unique holomorphic function $\overline\sigma = \overline\sigma(T, \mu)$ in $U$ satisfying \(F(\overline\sigma(T, \mu), T, \mu) = 0\).

% \cYL{discuss a little on its indications to Fig.11? e.g. the biholomorphic mapping means that the trajectories in Fig.11 does not intersect. This is is seen in the case when thermodynamic parameters $(T,\mu)$ do not reach the LYE singularity. ``Therefore, LYE does not satisfy point 3 ($\partial F \partial \sigma \neq 0$), ...''}

This behavior is precisely observed in Fig.~\ref{solutiontrajectory}\textbf{a} and Fig.~\ref{solutiontrajectory}\textbf{b}, where each stationary solution traces out a smooth, non-intersecting trajectory $\overline\sigma(T, \mu)$ as the temperature varies. The biholomorphic nature of the inverse mapping in Eq.~\eqref{Hinverse2} ensures local uniqueness, thereby prohibiting trajectory crossings.

However, this conclusion is contradicted by the behavior at the CEP (or more generally at the LYEs), where two distinct solutions—the Nambu and Wigner branches—coexist at $(T_l, \mu_l)$. As shown in Fig.~\ref{solutiontrajectory}\textbf{c}, in any neighborhood of $(T_l, \mu_l) = (91.29, 185.57)~\text{MeV}$, there exist two distinct solution branches $\overline\sigma_1(T,\mu)$ and $\overline\sigma_2(T,\mu)$ satisfying the gap equation. This multiplicity violates the local uniqueness guaranteed by the biholomorphic nature of the inverse mapping in Eq.~\eqref{Hinverse2}.
Therefore, the assumption Eq.~\eqref{eq:nondegenerate-assumption} does not hold, and it follows that
\begin{equation}
    \left. \frac{\partial F}{\partial \overline\sigma} \right|_{(\overline\sigma_l, T_l, \mu_l)} = 0.
\end{equation}

\qed

\begin{figure}[tbp]
    \centering
    \includegraphics[width=0.75\linewidth]{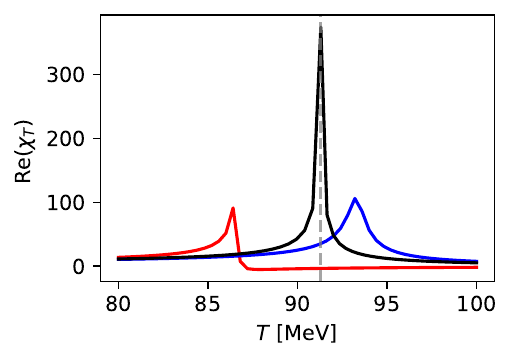}
    \caption{Temperature dependence of the real part of the susceptibility $\operatorname{Re}(\chi_T)$ for the positive Nambu solution, with $T$ ranging from 80 to 100~MeV at $\mu_I/\pi T = 0.001$, shown for several values of $\mu_R$. The red, blue, and black lines correspond to $\mu_R = 200.00$~MeV, $\mu_R = 180.00$~MeV, and $\mu_R = 185.57$~MeV, respectively. The black curve passes through the LYE, where a clear divergence in $\chi_T$ is observed.}
    \label{chimui001}
\end{figure}

As indicated by \Eqref{chiexp}, the susceptibility $\chi$ diverges at the LYE (or CEP) due to the coalescence of distinct solutions. This behavior is consistent with the results shown in Figure~\ref{fig:chimui0}. Figure~\ref{chimui001} illustrates the temperature dependence of $\operatorname{Re}(\chi_T)$ for the positive Nambu solution, with $T$ varying from 80 to 100~MeV at fixed $\mu_R = 185.57$~MeV and $\mu_I/\pi T = 0.001$. A clear divergence is observed at $T = 91.29$~MeV, signaling the location of the LYEs.

\begin{figure}[htbp]
    \centering
    \includegraphics[width=0.8\columnwidth]{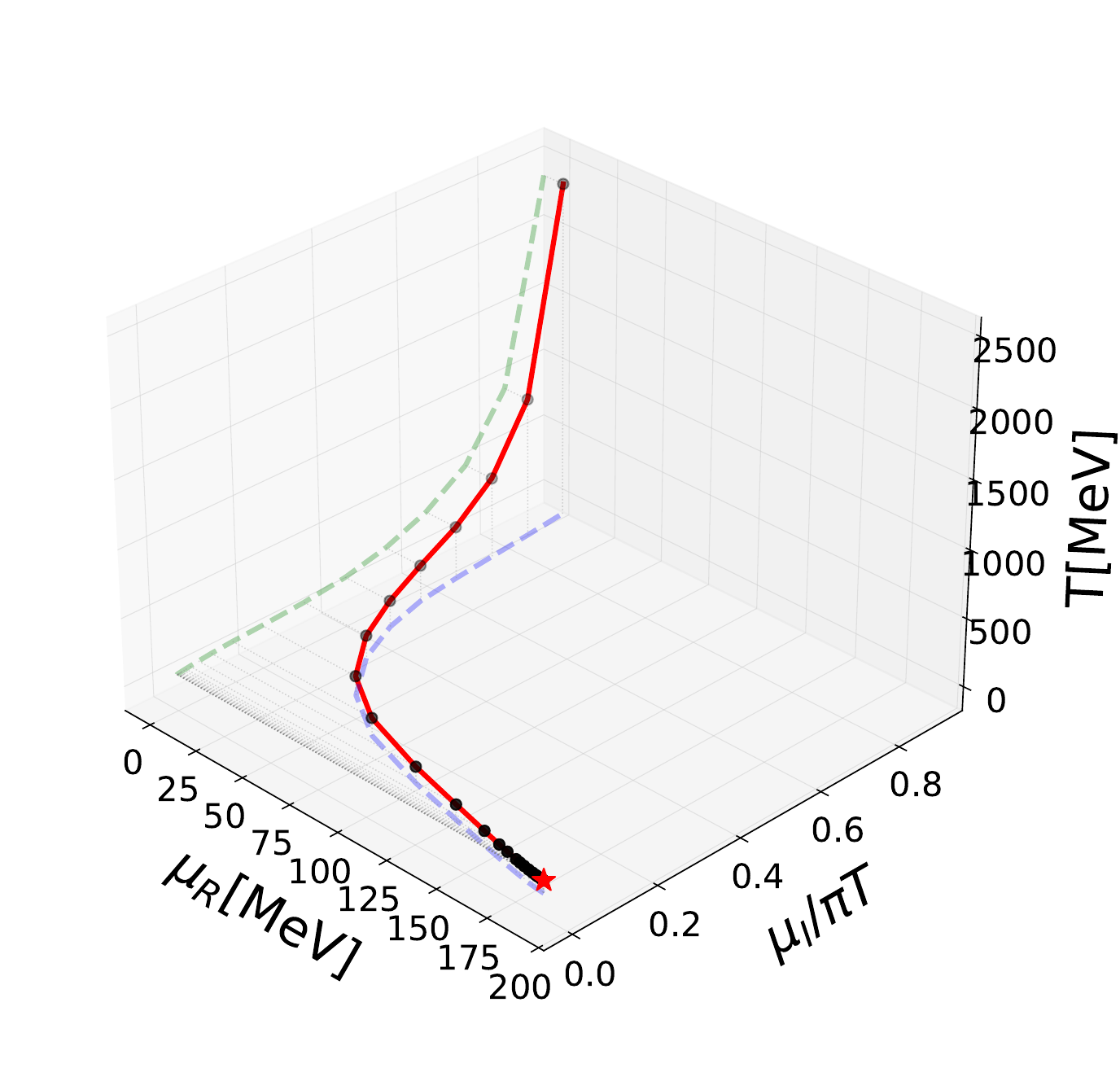}
    \caption{Trajectory of Lee--Yang edge singularities (LYEs) in the complex chemical potential plane for $\mu_I/\pi T \in [0,1]$.}
    \label{LYEs}
\end{figure}

% The thermodynamic potential $\Omega$ exhibits periodicity in the imaginary chemical potential $\mu_I$, with period $2\pi T$, as implied by the underlying structure of the partition function. In addition, it satisfies the reflection symmetry:
% \[
% \Omega(T, \mu_R, \mu_I, \overline\sigma) = \Omega(T, \mu_R, -\mu_I, \overline\sigma).
% \]
% Therefore, it suffices to consider only the region $\mu_I/\pi T \in [0,1]$. 
The resulting trajectories of Lee--Yang edge (LYE) singularities in the complex chemical potential plane are shown in Fig.~\ref{LYEs}.

The correlation length tends to diverge near the critical end point (CEP), and the associated critical behavior is governed solely by the symmetry and dimensionality of the system. Previous studies have shown that the trajectory of the Lee--Yang edge singularities (LYEs) near the CEP can be described by the scaling behavior of the Ising universality class~\cite{basarUniversalityLeeYangSingularities2021,wanLeeYangEdgeSingularities2024,clarkeDeterminationLeeYangEdge2023}:

\begin{equation}
    \mu_{\mathrm{LY}}(T) - \mu_{\mathrm{CEP}} = -c_1 (T - T_{\mathrm{CEP}}) + i c_2 (T - T_{\mathrm{CEP}})^{\beta \delta},
    \label{LYESextrapolation}
\end{equation}
where $c_1$ and $c_2$ are real-valued fitting coefficients.

Table~\ref{fittable} summarizes the fitting results for the distribution of LYEs based on the scaling relation above. The coefficient of determination $R^2$ represents the square of the Pearson correlation coefficient and quantifies the quality of the fit.

\begin{figure}[tbp]
    \centering
    \captionsetup{justification=centering}
    \includegraphics[width=0.8\linewidth]{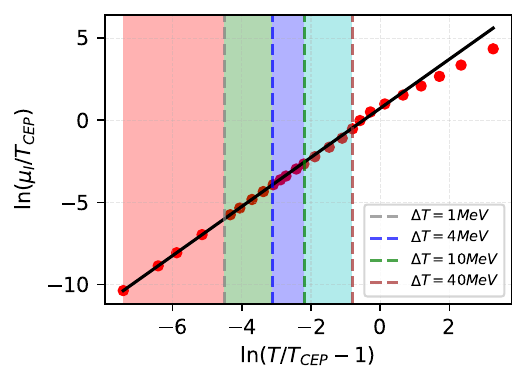}
    \caption{Scaling behavior of the LYEs.}
    \label{LYEsfit}
\end{figure}

By requiring a precision of three decimal places for $\beta\delta$ and $R^2 > 0.999999$, the critical region is found to be approximately $\Delta T \sim 1~\text{MeV}$ and $\Delta \mu_R \sim 3~\text{MeV}$. In this region, the extracted critical exponent in the non-local NJL model is $\beta\delta = 1.494(1)$, which is in close agreement with the mean-field values $\beta^{\mathrm{MF}} = 0.5$ and $\delta^{\mathrm{MF}} = 3.0$~\cite{PhysRevC.58.1758,dummCharacteristicsChiralPhase2005}.
Relaxing the constraint on the coefficient of determination to $R^2 > 0.99999$, the critical region broadens to approximately $\Delta T \sim 4~\text{MeV}$ and $\Delta \mu_R \sim 10~\text{MeV}$. The corresponding fit is illustrated in Fig.~\ref{LYEsfit}.

\begin{table}[tbp]
\centering
\caption{Fitting results for the critical scaling behavior of LYEs in different intervals. Here, $R^2$ denotes the coefficient of determination, and $\beta\delta$ is the extracted critical exponent for each interval. The quantities $\Delta T = T - T_{\mathrm{CEP}}$ and $\Delta \mu_R = \mu_R - \mu_{R,\mathrm{CEP}}$ are given in MeV. Assuming $\beta\delta = 1.5$, the location of the CEP is determined by fitting the LYEs trajectory within each interval.}
\begin{tabular}{ccccc}
\toprule
               & Interval1        & Interval2        & Interval3        & Interval4        \\
\midrule
$\Delta T$     & [0.05, 1.18]     & [1.18, 4.09]     & [4.09, 9.94]     & [9.94, 40.46]    \\
$\Delta \mu_R$ & [0.14, 3.17]     & [3.17, 10.80]    & [10.80, 25.88]   & [25.88, 114.30]  \\
$\Delta \frac{\mu_I}{\pi T}$ & [0.00001, 0.001] & [0.001, 0.006]   & [0.006, 0.02]    & [0.02, 0.13]     \\
$\beta\delta$  & 1.494(1)         & 1.470(2)         & 1.425(4)         & 1.50(5)          \\
$R^2$          & 0.999998         & 0.999990         & 0.999973         & 0.996359         \\
fitCEP         & [90.10, 188.74]  & [90.05, 188.82]  & [89.73, 189.41]  & [91.64, 188.50]  \\
\bottomrule
\end{tabular}

\label{fittable}
\end{table}

It has been mentioned that lattice QCD faces significant challenges at non-zero quark chemical potential $\mu_R$. Nevertheless, when $\mu_R/T$ is sufficiently small, the positions of Lee--Yang edge singularities (LYEs) can still be identified. Consequently, the critical end point (CEP) can be determined by extrapolating the trajectory of LYEs using Eq.~\eqref{LYESextrapolation}.

\begin{figure}[tbp]
    \centering
    \includegraphics[width=0.8\linewidth]{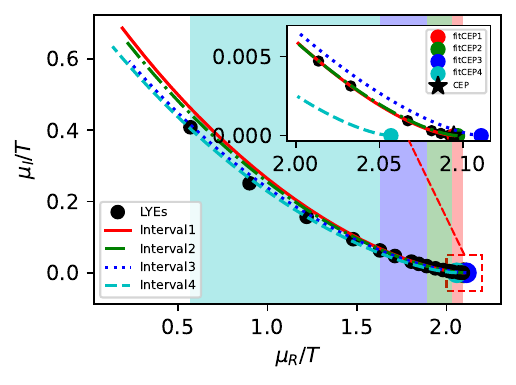}
    \caption{Determining the CEP via extrapolation of LYEs in several interval.}
    \label{LYEsfitextrapolation}
\end{figure}

To assess the validity of this method, we perform fits in several intervals, as shown in Fig.~\ref{LYEsfitextrapolation}, assuming a critical exponent of $\beta\delta = 1.5$. When the fitting region satisfies $\Delta T < 4~\text{MeV}$, $\Delta \mu_R < 10~\text{MeV}$, and $\Delta (\mu_I/\pi T) < 0.001$, the extrapolated CEP position closely matches the true value $T_{\mathrm{CEP}}, \mu_{\mathrm{CEP}} = (90.10, 188.73)~\text{MeV}$, with an error less than $0.1~\text{MeV}$.

However, as $\mu_R/T$ becomes smaller, the extrapolation based on Eq.~\eqref{LYESextrapolation} becomes less reliable. In Interval 3, the extracted critical exponent $\beta\delta$ is lower than expected, while in Interval 4, the coefficient of determination is significantly reduced. Despite these deviations, the estimated location of the CEP remains reasonably accurate.

In short, the extrapolation method based on analyzing LYEs proves to be effective for locating the CEP, provided that the positions of the LYEs can be determined with sufficient high precision.


\section{Summary}\label{sec:conclusion}
In this work, we have investigated the QCD phase diagram in the two-flavor non-local Nambu--Jona-Lasinio(NJL) model. The model introduces a momentum dependent form factor which characterizes the dynamical properties of strong interaction for the combined effects from gluon propagator and quark-gluon vertex. 
Using model parameters fixed in the vacuum as given in Refs.~\cite{dummStronginteractionMatterExtreme2021}, the chiral phase transition line is studied, with its curvature found to be $\kappa = 0.01708(2)$, which is in good agreement with previous estimation. 
In addition, the critical endpoint (CEP) is found at $T = 90.10\,\text{MeV}$ and $\mu = 188.73\,\text{MeV}$. 
%This consistency confirms the capability of the non-local NJL model to effectively interpolate between the results at real and imaginary chemical potentials.

%Furthermore, both the chemical potential $\mu$ and the chiral order parameter $\overline\sigma$ have been analytically continued to the complex plane. 
This further allows for the investigation of QCD phase transition in the presence of complex chemical potential, where the chiral order parameter $\overline\sigma$ is also extended to be a complex one. 
We have proposed an improved method for identifying the Lee--Yang edge singularities (LYEs), based on the classification of the potential $\Omega$ according to its shape. The phase diagram is divided into several regions characterized by distinct potential shapes, and the boundaries between them are referred to as ``shape-shifting lines". These lines intersect at the CEP or more generally at LYEs.

At LYEs, the positive Nambu solution and the Wigner solution always coalesce, implying that $\partial^2\Omega/\partial\overline\sigma^2 = 0$ and a divergence in the susceptibility. 
This behavior is further demonstrated in the trajectories of the gap equation solutions, where the LYEs corresponds to the intersection point of these distinct solution branches.

By employing this method, the trajectory of the LYEs has been determined. The extracted critical exponent $\beta\delta$ is 1.494(1) in the region $\Delta T \sim 1~\text{MeV}$ and $\Delta \mu_R \sim 3~\text{MeV}$, which is consistent with the mean-field theoretical prediction $\beta\delta = 1.5$. 
% \cYL{some discussion on ``the critical region is small''.}

Finally, we highlight the practical significance of the LYEs-based extrapolation method for locating the CEP. While lattice QCD is known to suffer from severe limitations at nonzero real chemical potential, our analysis demonstrates that the trajectory of the LYEs, which can be accessed when $\mu_R/T$ is small, provides a viable path forward. By performing critical scaling fits in appropriate parameter intervals and assuming a mean-field critical exponent $\beta\delta = 1.5$, we show that the extrapolated position of the CEP can be determined with high precision—within $0.1~\text{MeV}$ of the true value—when the fitting region satisfies $\Delta T < 4~\text{MeV}$, $\Delta \mu_R < 10~\text{MeV}$, and $\Delta (\mu_I/\pi T) < 0.001$. 
Although the reliability of the extrapolation decreases as the fitting region moves further from the CEP (i.e., as $\mu_R/T$ becomes smaller), the deviations remain moderate. 
This suggests that, given sufficiently accurate determination of LYEs, the extrapolation method offers a robust and complementary approach to identifying the CEP in the QCD phase diagram.

% The smallness of the extracted critical region has important implications for both theoretical studies and experimental searches for the CEP. On the experimental side, signals of criticality—such as enhanced fluctuations in conserved charges—can only be observed if the system's trajectory in the $(T,\mu)$ plane passes sufficiently close to the critical region. A narrow critical region, therefore, poses significant challenges for locating the CEP in heavy-ion collision experiments. 
% From the theoretical perspective, this narrow window requires high precision in model calculations and careful fitting procedures, such as those used in lattice QCD extrapolations or functional approaches, to resolve the universal critical behavior. 
% The agreement of our extracted exponent with the mean-field prediction suggests that the nonlocal NJL model captures the essential physics in this confined region.

% In future work, we plan to include higher-order corrections beyond the mean-field approximation in \ref{eq:bosonize} \cite{hellDynamicsThermodynamicsNonlocal2009} to further explore the influence of nonlocal interactions on the critical behavior of the system.

\section{Acknowledgments}
We thank Zi-yan Wan and Hui-wen Zheng for valuable discussions.
This work was supported by the National Natural Science Foundation of China under Grants No.~12247107 and No.~12175007.

\bibliography{nnjl}

\end{document}